# A reworked SOBI algorithm based on SCHUR Decomposition for EEG data processing

Kalogiannis Gregory
Aristotle University of Thessaloniki
Department of Electrical and
Computer Engineering
Thessaloniki, Greece
gkalogiannis@ece.auth.gr

Karampelas Nikolaos
Aristotle University of Thessaloniki
Department of Electrical and
Computer Engineering
Thessaloniki, Greece
nkarampe@ece.auth.gr

Hassapis George
Aristotle University of Thessaloniki
Department of Electrical and
Computer Engineering
Thessaloniki, Greece
chasapis@eng.auth.gr

*Abstract* — In brain machine interfaces (BMI) that are used to control motor rehabilitation devices there is the need to process the monitored brain signals with the purpose of recognizing patient's intentions to move his hands or limbs and reject artifact and noise superimposed on these signals. This kind of processing has to take place within time limits imposed by the on-line control requirements of such devices. A widely-used algorithm is the Second Order Blind Identification (SOBI) independent component analysis (ICA) algorithm. This algorithm, however, presents long processing time and therefor it not suitable for use in the brain-based control of rehabilitation devices. A rework of this algorithm that is presented in this paper and based on SCHUR decomposition results to significantly reduced processing time. This new algorithm is quite appropriate for use in brain-based control of rehabilitation devices.

## I. INTRODUCTION

Motor rehabilitation devices, such as continuous passive motion machines (CPM), are used for rehabilitation in hospitals, clinics or general practices and they are important supplement to medical and therapeutic treatment. Their mode of operation is to move injured joint over a range of motion in a circular periodical way defined by the physician [1]. For example, in case of elbow and fist joints, these devices impose movement via flexion/extension and/or pronation/supination to the injured joint [2]. Although their indisputable contribution to rehabilitation [3], it is believed that the overall treatment time can be reduced and the overall rehabilitation could be improved if the patient interacts with these devices and their motion is determined according to patient's will.

These motor rehabilitation devices allow their easily connection with controllers which can make the devices to follow trajectories determined by processing the generated by the patient EEG brain signals in order to extract the patient's intensions and will. The implementation of such an architecture requires fast recognition of the motor imagery movements of the joint in order to create the appropriate control signals. This can be done by manipulating the EEG data with the purpose of removing the noise and information that is not essential for creating the control signal in a fast and effective way.

Independent component analysis (ICA) [4] is a special case of blind source separation and it is used widely not only for studying specific brain EEG activity and separating it from other non-brain activities [5] but also for separating artifacts from EEG data. ICA separates data from multi-channel signals the time courses of which are maximally independent from each other. The most common ICA algorithms used in EEG data analysis are Infomax ICA [6-7], SOBI [8], and FastICA [9]. All ICA algorithms have the same overall goal [10], and generally all of them produce near-identical results when applied to idealized (model) source mixtures. However, since EEG brain and non-brain source signals are not totally independent, sometimes different ICA algorithms may return slightly different results when applied to the same EEG data.

In case of a BMI that process EEG data with the purpose to control motor imagery movements, the overall process time becomes critical. Therefore, Blind Source Separation (BSS) [11] algorithms must be fast enough to perform effective EEG decomposition [12]. So, they must run within the time limits imposed by the movement control requirements.

In this paper, we demonstrate a new blind source separation technique based on SOBI algorithm. Reworking and substituting several steps of SOBI algorithm using Schur decomposition, we managed to improve the execution time of signal decomposition procedure. It is demonstrated experimentally that a significant computation time reduction is achieved using the rework SOBI against the typical EEG data decomposing technique.

In section II the problem formulation is presented while basic assumptions of EEG signal and Schur decomposition are stated. Current SOBI algorithm steps are explained and the mentioned above proposed rework of SOBI technique is presented. In Section III the rework technique is tested with pre-recorded EEG data obtained from signals corresponding to motor imagery movements and overall conclusions are drawn which are presented in the Section IV. Section V proposes future work that can be undertaken towards further reduction of algorithm execution time.

## II. PROBLEM FORMULATION

### A. Assumptions

The basic ICA model that SOBI is based on is expressed as

$$x(t) = g(t) + n(t) = As(t) + n(t) \quad (1)$$

where $s$ denotes the source signals, $x$ the signals we receive, $n$ the noise factor and $A$ is a mixing matrix. It is assumed that the source signal vector $s(t)$ is either a deterministic ergodic sequence or a stationary multivariate process. In such a case, the autocovariance function is

$$E[s(t+\tau)s(t)^*] = diag[\rho_1(\tau),...,\rho_n(\tau)] \quad (2)$$

where superscript * denotes the conjugate transpose of a vector, and *diag* is the diagonal matrix formed with the elements of its vector valued argument

$$\rho_i(\tau) = E[s_i(t+\tau) s_i(t)^*] \quad (3)$$

The noise that exists in this algorithm is modeled as a stationary, temporally white, zero mean complex random process that is independent of the source signals. For simplicity, we also require to be spatially white, i.e.,

$$E[n(t+\tau)n(t)^*] = \sigma^2 \delta(\tau) I \quad (4)$$

where $\delta(\tau)$ is the Kronecker delta, and denotes the identity matrix. Under the above assumptions, the covariance matrices of the array output take the following structure:

$$R(0) = E[x(t)x(t)^*] = AR_s(0)A^H + \sigma^2 I \quad (5)$$

$$R(\tau) = E[x(t+\tau)x(t)^*] = AR_s(\tau)A^H, \quad \tau \neq 0 \quad (6)$$

The basic goal of every blind source separation algorithm is to identify the mixture matrix A and the source signals without any previous knowledge of the array manifold. This method guarantees that the source separation is unaffected by errors in the reproduction model or in array calibration.

### B. SCHUR decomposition

SCHUR decomposition is a mathematical model that has been widely applied in many scientific areas such as Lie theory [13]. In this study, it is used to make important transformations to the blind source separation algorithm SOBI. According to SCHUR decomposition if $A$ is an $nxn$ square matrix with complex entries, it can be expressed as:

$$A = QUQ^{-1} \quad (7)$$

where $Q$ is a unitary matrix and $U$ is upper triangular. $U$ is called a SCHUR form of $A$. $U$ is similar to $A$ (two matrices $A$ and $U$ are similar if $U = PAP^{-1}$ for some invertible $nxn$ matrix $P$) so it has the same set of eigenvalues. Since it is triangular those eigenvalues are the diagonal entries of $U$ (since for any triangular matrix $A$ the matrix $\lambda I - A$, whose determinant is the characteristic polynomial of $A$, is also triangular, the diagonal entries of $A$ give the multiset of its eigenvalues. An eigenvalue with multiplicity k occurs k times as a diagonal entry). This kind of decomposition is known to be computed by the *QR* algorithm or its variants. [14]

### C. SOBI Algorithm

The algorithm works with the use of joint diagonalization on a set of partial covariance matrices. Other mathematical tools like whitening and theorems on unitary matrices are used as well but the basic concept is joint diagonalization. A basic assumption is that the signal vector $s(t)$ is a multivariate process. SOBI algorithm consists of the following steps:

*1)* Provided that Matrix $R(0)$ has $m$ eigenvalues, find the $n$ largest data samples largest from $a$ number of $T$ data samples. Then compute the corresponding eigenvectors.

*2)* Since the noise factor is temporally white, the noise variance will be estimated as the average of the rest $m-n$ eigenvalues.

*3)* The next step, called whitening, is basic and covers a big part of SOBI. This is used to transform the matrix $x(t)$ (which is basically a vector of random variables with a known covariance matrix) into a matrix, of a new set of variables, whose covariance is the identity matrix. Practically, whitening changes the input vector into a white noise vector. The whitening matrix used in SOBI is computed as follows:

   *a)* The whitening signals $g(t)=[g_1(t),...,g_n(t)]^T$ will be computed by the formula $g_i(t) = (\lambda_i - \sigma^2)^{-(1/2)} h_i^* x(t)$ for $1 \leq i \leq n$. Equivalently, the whitening matrix is formed as:

$$W = [(\lambda_1 - \sigma^2)^{-(1/2)} h_1, ..., (\lambda_n - \sigma^2)^{-(1/2)} h_n]^H \quad (8)$$

   *b)* Since $W$ is a whitening matrix, $WA$ is a $nxn$ unitary matrix such that $WA = U$. So, $A$ can be factored as $A = W^\# U$ where superscript # denotes the Moore-Penrose pseudo-inverse. The fact that $WA$ is unitary arises as follows: In order to whiten $g(t)$ we apply to it a whitening matrix $W$.

$$E[Wg(t)g(t)^* W^H] = WR_g(0)W^H = WAA^H W^H = I \quad (9)$$

From (8) we obtain that $WA(WA)^H = I$ which proves that $WA$ is unitary. The whitening of the data in SOBI is done directly using Singular Value Decomposition (SVD) [15-16].

*4)* Compute sample covariance matrices $R(\tau)$ of $g(t)$ for a set of time lags $\tau \in \{ \tau_j \mid j=1,...,L\}$ that is fixed.

*5)* The "off" of an $nxn$ matrix $M$ with entries $M_{ij}$ is defined as the sum of all the entries, squared, that do not belong to the diagonal. Mathematically this is expressed as

$$\text{off}(M) := \sum_{1 \leq i \neq j \leq n} |M_{ij}^2| \quad (10)$$

To diagonalize a matrix $M$ unitarily is equivalent to zeroing $\text{off}(V^H MV)$. $V$ must be unitary. The key point of this algorithm is that if a matrix can be written in the form

$$M = UDU^H \quad (11)$$

where $D$ is diagonal and $U$ is unitary with district diagonal elements, then it can be unitarily diagonalized. In addition, it can be unitarily diagonalized only by unitary matrices that are essentially equal to $U$. So, if $\text{off}(V^H MV)=0$ then $V$ is essentially equal to $U$.

If a matrix is normal ($MM^H=M^HM$) it is unitarily diagonalizable. That is equal to the existence of a unitary matrix $U$ and a diagonal matrix $D$ such that (11) is true for this matrix.

For a set of matrices e.g. $R(\tau)$, $\tau \in \{\tau_j | j=1,...,L\}$ a joint diagonalizer is a unitary matrix that minimizes the sum of all the off functions of the sum: $\text{off}(VRV^H)$. If each of these matrices $R(\tau)$ can be written in the form $R(\tau)= UDU^H$ then obviously $V = U$ and the sum (10) is equal to zero. $U$ is called the joint diagonalizer of the set. In this step of the algorithm a unitary matrix U is used as joint diagonalizer for the set $\{R(\tau_j) | j=1,...,L\}$.

6) The source signals can now be estimated as s(t) = UHWx(t) and the mixing matrix A is computed as A = W # U.

### D. SOBI using SCHUR Implementation

The part of SOBI that is going to be modified in this paper is the way $\{R(\tau_j) | j=1,...,L\}$ is diagonalized. In general, any whitened covariance matrix is diagonalized by the unitary transform $U$. [17] As mentioned above obtaining the unitary factor $U$ is equal to obtaining a unitary diagonalizing matrix of the whitened $\{R(\tau_j) | j=1,...,L\}$. SOBI obtains $U$ as a joint diagonalizer of the set $\{R(\tau_j) | j=1,...,L\}$ using Given's rotation. [18-19]

Given's rotation is a method used to diagonalize nxn matrices though multiplying them with matrices that are computed with a specific formula [19]. Practically it is a rotation in the plane spanned by two cooedinates axes. In every step of this method an entry is zeroed until the array is diagonalized. The basic con of this formula is that most of the times a large set of matrices has to be created(one matrix for every entry that has to be zeroed), which makes the algorithm slower and consumes a lot of memory. Especially in EEG experiments where the amount of the observations is large (correlation matrices are large) Given's rotation is a slow process.

Another way to execute SOBI is through a variant of the QR algorithm. QR is an eigenvalue algorithm based on SCHUR decomposition. The basic advantage of SOBI using SCHUR decomposition is that this way no external matrices (like the matrices used in Given's Rotation) have to be computed. The only matrix this algorithm works with is the matrix that has to be diagonalized This method saves time and a lot of memory as well. Practically, it calculates eigenvalues and eigenvectors of a matrix. The idea is to write a matrix as a product of an orthogonal matrix and an upper triangular. In this algorithm, the diagonalization is computed as follows:

Samples of $\{R(\tau_j) | j=1,...,L\}$ have already been formed on the previous step of SOBI. Generally, if a unitary factor U diagonalizes one of these samples it diagonalizes the rest as well. The first step of the new algorithm is computing the SCHUR form of the first sample. That means that $R(\tau_1)$ is equal to

$$R(\tau_1) = QBQ^H \quad (12)$$

with $Q$ being a unitary matrix and $B$ an upper triangular.

A very crucial part of this algorithm is noticing that since $B$ and $R(\tau_1)$ are similar ($R(\tau_1)=P^{-1}BP$, for some invertible P) they have the same multiset of eigenvalues and since $B$ is triangular those eigenvalues are the diagonal entries of $B$. Since all of the matrices $R(\tau_j)$ have a set of district eigenvalues, $B$ has a set of district entries on its diagonal. This can be shown as follows:

$$(b_{ii} - \lambda) \quad (13)$$

With $b_{ii} \neq b_{jj}$ for $i \neq j$ so every eigenvalue has multiplicity 1. So, $B$ is diagonalizable. Matrix $B$ can be written as $B=VDV^{-1}$ (V is unitary so $V^{-1} = V^H$) and $R(\tau_1) = QVDV^HQ^H = (QV)D(QV)^H$. The unitary matrix that can diagonalize the matrix $R(\tau_1)$ is unique. This is the reason why $V^{-1} = V^H$.

Now the matrix $R(\tau_1)$ is diagonalized. As it was explained any matrix that diagonalizes $R(\tau_1)$ is essentially equal to the unitary factor $U$ from (12) so we obtain that $U=QV$. The final step of this algorithm is to compute the source signals and the mixing matrix $A$ from equation (1). The source signals are estimated as $s(t)=U^HWx(t)$ and the mixing matrix A as $A=W^{\#}U$.

### III. EXPERIMENTAL RESULTS

In order to evaluate the computational time and effectiveness of the proposed reworked SOBI algorithm, several experiments were conducted using prerecorded EEG datasets. These prerecorded datasets were created and contributed to PhysioNet [20] database by the developers of the BCI2000 instrumentation system [21]. Datasets includes different sessions of over 1500 of one and two-minute EEG recordings that correspond to 109 volunteers. The volunteers performed different motor/imagery tasks while 64-channel EEGs were recorded. Over 90% of randomly selected sessions from the above database where used, where each session contains 14 recordings of EEG data. For each session and recording, the execution time of the reworked SOBI algorithm was compared with the execution time of the typical SOBI approach. Experiments were conducted using EEGLAB Toolbox [22] for MATLAB, run on an Intel Core i7 – 2600K at 3.70GHz machine, with 16 GB of RAM.

In all cases, reworked SOBI presents a reduction in execution time comparing to execution time of the typical SOBI algorithm. Table I presents respectively the five best results while Table II the five worst results in our experiments.

TABLE I. EXECUTION TIME OF REWORKED AND TYPICAL SOBI BLIND SOURCE SEPARATION ALGORITHM, IN SECONDS – FIVE BEST RESULTS

| Best Result | Execution Time in sec | | Time Reduction in sec |
|---|---|---|---|
| | *Rework SOBI* | *Typical SOBI* | |
| 1 | *1.84* | *11.66* | *9.82* |
| 2 | *1.81* | *11.51* | *9.70* |
| 3 | *1.84* | *10.37* | *8.53* |
| 4 | *1.60* | *9.79* | *8.19* |
| 5 | *1.73* | *9.80* | *8.07* |

TABLE II.   EXECUTION TIME OF REWORKED AND TYPICAL SOBI BLIND SOURCE SEPARATION ALGORITHM, IN SECONDS – FIVE WORST RESULTS

| Worst Result | Execution Time in sec | | Time Reduction in sec |
|---|---|---|---|
| | *Rework SOBI* | *Typical SOBI* | |
| 1 | *5.02* | *9.33* | *4.31* |
| 2 | *5.22* | *9.87* | *4.65* |
| 3 | *5.69* | *10.52* | *4.83* |
| 4 | *4.78* | *10.22* | *5.44* |
| 5 | *4.45* | *11.12* | *6.67* |

## IV. CONCLUSIONS

This paper introduces an improved variant of SOBI algorithm, which is used as an ICA filter during brain signal processing. It is based on the joint diagonalization of a capricious set of covariance matrices and SCHUR decomposition. It allows the separation of Gaussian sources as the original SOBI algorithm does. However, compared to the original SOBI algorithm it presents a significantly lower processing time, robustness and less memory consumption. In specific, the reduction in the processing time is well below the limits imposed by the control of the motor imagery movements, a feature that makes this algorithm quite appropriate for use in the control of arm or limb rehabilitation devices.

## V. FUTURE WORK

Although the most time-consuming part of the algorithm processing is the diagonalization, there are still many other parts that can be modified in order to reduce further he execution time. The basic changes that can be made are:

*1)* The number of correlation matrices to be diagonalized. If this number can be further geuced, without having the source signals differentiated in the end, the algorithm will become faster. This is obvious since the number of the matrices to be processed will be reduced.

*2)* There are many ways of whitening data in signal processing. Although SVD is a basic formula it can get quite complex. The definition of a less complex whitening matrix from (8) could reduce the processing time.